\documentclass[11pt]{article}
\pdfoutput=1
\usepackage[latin1]{inputenc}
\usepackage{amsmath}
\usepackage{amsfonts}
\usepackage{amssymb}
\usepackage{amsthm}
\usepackage{mathrsfs}
\usepackage{amssymb}
\usepackage{cancel}
\usepackage{slashed}
\usepackage{graphicx}
\usepackage[font=small,labelfont=bf]{caption}
\usepackage{float}
\usepackage{csquotes}
\usepackage[bottom]{footmisc} %puts the footnotes in the proper place
\usepackage{tikz}
\usepackage{setspace}
\usepackage{bbding}
\usepackage{cite}
\usepackage{array,longtable}
\makeatletter

\@addtoreset{equation}{section}
\makeatother

\def\lsim{\;\raise0.3ex\hbox{$<$\kern-0.75em\raise-1.1ex\hbox{$\sim$}}\;}
\def\gsim{\;\raise0.3ex\hbox{$>$\kern-0.75em\raise-1.1ex\hbox{$\sim$}}\;}
\def\beq{\begin{equation}}   \def\eeq{\end{equation}}
\def\ba{\begin{array}}       \def\ea{\end{array}}
\def\bea{\begin{eqnarray}}   \def\eea{\end{eqnarray}}
\def\nn{\nonumber}
\def\nl{\newline}

\theoremstyle{definition} 
\date{\today}

\begin{document}

\begin{titlepage}
%\begin{flushright}
%LPT Orsay 19-XX
%\end{flushright}

%\centerline{\bf \today}

\begin{center}

%\begin{doublespace}

\vspace{1cm}
{\Large\bf Weyl Consistency Conditions from a local Wilsonian Cutoff} 
\vspace{2cm}

{\bf{Ulrich Ellwanger}}\\
\vspace{1cm}
{\it  University  Paris-Saclay,  CNRS/IN2P3,  IJCLab,  91405  Orsay,  France}

%\end{doublespace}
\end{center}

\vspace*{2cm}
\begin{abstract}

A local UV cutoff $\Lambda(x)$ transforming under Weyl rescalings allows to construct Weyl invariant kinetic terms for scalar fields including Wilsonian cutoff functions. First we consider scalar fields in curved space-time with local bare couplings of any canonical dimension, and anomalous dimensions which describe their dependence on the UV cutoff. The local component of the UV cutoff plays the role of an additional coupling, albeit with a trivial constant $\beta$ function. This approach allows to derive Weyl consistency conditions for the corresponding anomalous dimensions which assume the form of an exact gradient flow. For renormalizable theories the Weyl consistency conditions are initially of the form of an approximate gradient flow for the $\beta$ functions, and we derive conditions under which it becomes the form of an exact gradient flow.

\end{abstract}

\end{titlepage}

\section{Introduction}

Weyl consistency conditions have lead to remarkable insights into quantum field theories. In the form derived by Osborn and Jack and Osborn in \cite{Osborn:1989td,Jack:1990eb,Osborn:1991gm} [JO] the Weyl consistency conditions imply a gradient flow for the renormalization group flow described by $\beta$~functions in low orders in perturbation theory, a phenomenon observed earlier in \cite{Wallace:1974dx,Wallace:1974dy}. One motivation was to derive a $c$~(or $a$)~theorem in four dimensions, but of interest are also the induced relations among coefficients of $\beta$~functions or anomalous dimensions of composite operators, and the ultraviolet (UV) and infrared (IR) asymptotic behaviour of quantum field theories.

The derivation of Weyl consistency conditions requires to consider a quantum field theory in curved space-time described by a background metric $\gamma_{\mu\nu}$, and to promote couplings $g_i$ to local couplings $g_i(x)$. Under local Weyl rescalings $\delta_\sigma \gamma_{\mu\nu}= -2\sigma \gamma_{\mu\nu}$, the local couplings transform according to their anomalous dimensions or $\beta$~functions. The Weyl consistency conditions in the form derived in [JO] follow either from the fact that local Weyl rescalings (being abelian) acting on the vacuum partition function via Weyl rescalings of $\gamma_{\mu\nu}$ and/or via Weyl rescalings of the local couplings have to commute, or from the finiteness of $\beta$~functions in dimensional regularization.

Implied relations among coefficients of $\beta$~functions in dimensional regularization have been studied in \cite{Antipin:2013sga,Jack:2013sha,Jack:2014pua,Jack:2015tka,Antipin:2018brm,Poole:2019txl,Poole:2019kcm,Davies:2019onf,Sartore:2020pkk,Steudtner:2020tzo} for the Standard Model and others. Implications of Weyl consistency conditions for higher dimensional field theories have been investigated e.g. in \cite{Grinstein:2013cka,Gracey:2015fia}.
Actually the gradient flow property of the renormalization group flow in $d=4$ is not exact, a priori confined to low orders in perturbation theory. Its general validity was challenged e.g. in \cite{Fortin:2011sz}, see also the discussion in \cite{Baume:2014rla}.

By far most of the previous explicit calculations were performed using dimensional regularization where classical Weyl invariance is broken by the scale $\mu$, introduced for otherwise dimensionless couplings. Here we consider a regularization by a Wilsonian cutoff $\Lambda$, by which we understand a modification of kinetic terms in the action such that modes with momenta $p^2 \gg \Lambda^2$ are suppressed, i.e. the kinetic terms become very (e.g. exponentially) large for $p^2 \gg \Lambda^2$.

The implementation of a Wilsonian cutoff in an otherwise Weyl invariant theory has been considered in \cite{Codello:2012sn,Codello:2013iqa,Codello:2015ana}, with the aim to study exact (functional) renormalization group equations (RGEs) in 2 and 4 dimensions. The authors concluded that a cutoff $\Lambda$ has to become local, i.e. $\Lambda(x)$. However, the authors restricted the dependence of $\Lambda$ on $x$ to be given by a dilaton (a Weyl compensator) which has dynamics in principle, but can be gauged away if Weyl invariance remains unbroken. Local couplings have not been introduced, and Weyl consistency conditions have not been considered. Another approach to implement a Wilsonian cutoff in a Weyl invariant way has been proposed in \cite{Rosten:2018cyr}, again in the context of exact RGEs. A local renormalization group equation was considered as early as 1987 in \cite{Shore:1986hk} in two-dimensional curved spacetime sigma models in order to derive consistency conditions on allowed backgrounds in string theory. In \cite{Ciambelli:2019bzz} a local cutoff was introduced for the study of Weyl invariance in the framework of holography, and in \cite{Melo:2019zay} in the context of the quantum renormalization group.

Here we present a way to implement a Wilsonian cutoff respecting Weyl invariance, based on a local cutoff $\Lambda(x)$ which transforms under Weyl rescalings. As a consequency all parameters (couplings and masses) transform under Weyl rescalings according to their canonical plus anomalous dimensions, except for the renormalization scale $\mu$ itself. Besides $\mu$ all parameters, couplings and masses, are considered as local. This allows to deduce a local RGE, a notion used to define the response of the action including counter terms with respect to Weyl rescalings $\delta_\sigma$. It implies a local RGE satisfied by the vacuum partition function which depends on the background metric, local couplings and covariant derivatives thereof as described in [JO]. The local cutoff $\Lambda(x)$ appears in this RGE like a local coupling, albeit with a trivial constant ``$\beta$~function''. For renormalizable theories, dimensional analysis can be used to relate $\beta$~functions of couplings with respect to the UV cutoff to $\beta$~functions with respect to a renormalization scale $\mu$.

The purpose of this article is to derive Weyl consistency conditions, and conditions leading to gradient flows for $\beta$~functions or anomalous dimensions in the space of couplings enlarged by $\Lambda(x)$. We consider two distinct cases: Case~1 where a true physical cutoff exists, bare couplings vary with the cutoff, and one is interested in the variation of the vacuum partition function with the cutoff. Here the limit of an infinite cutoff may not exist. Case~2 corresponds to a renormalizable quantum field theory where the limit of an infinite cutoff exists provided a finite number of counter terms is added. We show that a gradient flow holds in the space of anomalous dimensions $\gamma_n$ of composite operators ${\cal O}_n$ enlarged by $\Lambda(x)$ in case~1, and under a certain condition in the space of $\beta$ functions in case~2.

In section~2 we introduce Weyl invariant Wilsonian cutoff functions which allow to derive local RGEs for case~1 and case~2 in section~3. In section~4 these are applied to the vacuum partition function. Weyl consistency conditions are derived for case~1 in Section~4.1, and for case~2 in Section~4.2.

As a very first application we consider a real scalar field $\varphi$ with mass $m$ and quartic coupling~$g$ in four dimensions in section~5. We compute the metric relevant for the gradient flow to lowest nontrivial order $g^1$, where only the $\beta$ functions in the subspace $m(x),\Lambda(x)$ contribute. To this end we employ a new method to compute this metric in momentum space, in a weak field expansion around flat space time and in fluctuations of local couplings (including $m(x),\Lambda(x)$) around constant values. The results for the simple scalar model allow for first consistency checks of the methods. 

In Appendix~A we describe how to compute the metric in momentum space, in Appendix~B we explain under which condition the component of the metric with one index corresponding to $\Lambda(x)$ has the form of a gradient in the space of the remaining couplings. In Appendix~C we compute vertices from the kinetic term including a cutoff function and describe the relevant diagrams.
A summary and conclusions are given in section~6.

\section{Weyl Invariant Wilsonian Cutoff Functions}

In this section we construct a Weyl invariant kinetic term, including a Wilsonian UV cutoff, for a real scalar field $\varphi$ in $d$ dimensions. The UV cutoff $\Lambda$ is assumed to be local, i.e. $\Lambda(x)$, and to transform under Weyl rescalings together with the background metric $\gamma_{\mu\nu}$ and the scalar field:
\bea
&\delta_\sigma \gamma_{\mu\nu}=& -2\sigma \gamma_{\mu\nu}\; ,%\qquad (\delta_\sigma \sqrt{\gamma}=-d\sigma \sqrt{\gamma})\; ,
\nn \\
&\delta_\sigma \varphi=&  \sigma \left(\frac{d}{2}-1\right) \varphi\; ,
\nn \\
%&\delta_\sigma m^2=& 2\sigma m^2
%\nn \\
&\delta_\sigma \Lambda=& \sigma \Lambda\; .
\label{eq:weylresc}
\eea

The construction of a kinetic term including a Wilsonian cutoff will proceed stepwise. We start with the known expression for a Weyl covariant generalization of the covariant Laplacian
\beq
\nabla^2-\xi R\; , \quad 
\nabla^2=\frac{1}{\sqrt{\gamma}}\partial_\mu\sqrt{\gamma}\gamma^{\mu\nu}\partial_\nu\; ,
\quad \gamma=\det({\gamma_{\mu\nu}})\; ,
\quad \xi=\frac{d-2}{4(d-1)}\; ,%\quad \left(\xi_{d=4}=\frac{1}{6}\right)
\eeq
which satisfies
\beq
\delta_\sigma\left[(\nabla^2-\xi R){\cal O}\right] = \left(\frac{d}{2}+1\right) \sigma \left[(\nabla^2-\xi R){\cal O}\right]
\eeq
provided the operator $\cal{O}$ satisfies $\delta_\sigma {\cal O}=\sigma \left(\frac{d}{2}-1\right) {\cal O}$ such that, for ${\cal O}=\varphi$, $\delta_\sigma (\sqrt{\gamma}\varphi(\nabla^2-\xi R)\varphi)=0$ (using $\delta_\sigma \sqrt{\gamma}=-d\,\sigma \sqrt{\gamma}$). A Weyl invariant Wilsonian cutoff can be constructed with help of the operator
\beq
D_\Lambda = \Lambda^{-2}(\nabla^2-\xi R)\; .
\eeq
Using $\delta_\sigma \Lambda^{-2}=-2\sigma \Lambda^{-2}$, any expression of the form 
\beq\label{eq:FD}
F(D_\Lambda)% = \Lambda^{-1}f(D_\Lambda)\Lambda \equiv f(\Lambda^{-2}(\nabla^2-\xi R))
\eeq
satisfies
\beq
\delta_\sigma \left( F(D_\Lambda) \varphi\right) = \left(\frac{d}{2}-1\right)\sigma  F(D_\Lambda) \varphi\; .
\eeq
The possibility to construct $F(D_\Lambda)$ with this property under local Weyl transformations requires the use of the $x$~dependent cutoff $\Lambda$ transforming as in eq.~\eqref{eq:weylresc}. An example is
\beq\label{eq:FDe}
F(D_\Lambda)=e^{-D_\Lambda}
\eeq
which leads to an exponentially suppressed propagator for large $p^2$ in momentum space.
The kinetic part $S_k$ of the action reads then
\beq\label{eq:kin}
S_k=\frac{1}{2}\int\sqrt{\gamma}d^dx\, \varphi(-\nabla^2+\xi R)\, F(D_\Lambda) \varphi
\eeq
and satisfies
\beq\label{eq:rge1}
\int\sqrt{\gamma}d^d x\, \sigma 
\left\{-2 \gamma_{\mu\nu}\frac{\delta}{\delta \gamma_{\mu\nu}}
+ \varphi \frac{\delta}{\delta \varphi}
+\Lambda \frac{\delta}{\delta \Lambda}
 \right\}  S_k=0\; .
\eeq

\section{The Action and Local Renormalization Group Equations}

As interactions we consider operators ${\cal O}_n(\varphi,\nabla) \equiv {\cal O}_n(\varphi,\gamma_{\mu\nu})$ which satisfy
\beq\label{eq:defop}
\left\{
\int \sqrt{\gamma} d^dy\, \sigma 
\left(-2 \gamma_{\mu\nu}\frac{\delta}{\delta \gamma_{\mu\nu}}
+ \varphi \frac{\delta}{\delta \varphi}\right)
 \right\}\sqrt{\gamma(x)}\, {\cal O}_n(x)= -d_n^0 \sigma(x) \sqrt{\gamma(x)}\, {\cal O}_n(x)\; .
\eeq
Typically one has ${\cal O}_n=\varphi^n$ such that $d_n^0 = d-n$. The interaction part $S_{int}$ involving real scalar fields $\varphi\equiv \left\{\varphi_k\right\}$ (indices of fields will be suppressed for simplicity) reads
\beq\label{eq:sint}
S_{int}=\int \sqrt{\gamma} d^dx \sum_n g_n^0(\Lambda)\, {\cal O}_n
\eeq
where $g_n^0(\Lambda)$ are bare marginal, relevant or irrelevant couplings of canonical dimension $d_n^0$.
A Weyl rescaling of a term $\sim \sqrt{\gamma}\, g_n^0(\Lambda)\, {\cal O}_n$ in $S_{int}$ gives
\beq\label{eq:rgen}
\left\{\int\sqrt{\gamma} d^d x\, \sigma 
\left(-2 \gamma_{\mu\nu}\frac{\delta}{\delta \gamma_{\mu\nu}}
+ \varphi \frac{\delta}{\delta \varphi}
+\Lambda \frac{\delta}{\delta \Lambda}\right)
 \right\}\sqrt{\gamma}\,  g_n^0\, {\cal O}_n=-d_n^0 g_n^0 \sigma \sqrt{\gamma}\, {\cal O}_n
+\sqrt{\gamma}\, {\cal O}_n \int\sqrt{\gamma} d^dx  \sigma \Lambda \frac{\delta g_n^0}{\delta \Lambda} 
\; .
\eeq
In the following we consider two different scenarios:

Case 1: One may consider $S_{int}$ as an effective action valid at scales $< {\Lambda}$ where momenta $> {\Lambda}$ have already been integrated out. Then $S_{int}$ can include irrelevant operators multiplied by couplings $g_n^0(\Lambda) \sim \Lambda^{d_n^0}$ with $d_n^0 < 0$. The aim will be to derive a local RGE including the dependence of the full action $S=S_k+S_{int}$ on the local cutoff $\Lambda$.

Case 2 corresponds to a renormalizable theory where $g_n^0$ are bare marginal or relevant couplings of canonical dimension $d_n^0$ with $0 \leq d_n^0 < d$. $g_n^0$ include $\Lambda$ and $\mu$ dependent counter terms such that Green functions of $\varphi$ are finite for $\Lambda \to \infty$. We derive a local RGE including the dependence on the local cutoff $\Lambda$ which is valid before the limit $\Lambda \to \infty$ is taken.
Subsequently we consider vacuum diagrams only, thus no sources need to be coupled to $\varphi$. Hence, apart from a trivial factor originating from the measure, the path integral allows for redefinitions of $\varphi$. This freedom is used to remove wave function renormalisation factors $Z$ multiplying the kinetic term. As a consequence of these field redefinitions the counter terms in $g_n^0$ include contributions from the $Z$ factors. (Since $Z$ factors depend on $\Lambda(x)$ and $g_i(x)$ they are local as well. In the case of several interacting scalars, the treatment of terms $\sim \partial_\mu Z(x)$ requires special care.)

In case 1 we assume that the $\Lambda$ dependent bare couplings $g_n^0$ satisfy scaling relations of the form 
\beq\label{eq:scalrel}
\Lambda\frac{\delta}{\delta \Lambda} g_n^0 = \hat{\gamma}_n(\Lambda, g_i^0)\, g_n^0 \; .
\eeq
(Throughout this paper $\delta$ denote functional local derivatives. Appropriate factors of $\int \sqrt{\gamma}$ are understood.)
Then the full action $S=S_k+S_{int}$ satisfies a local RGE
\beq\label{eq:rgen1}
\int\sqrt{\gamma}d^d x\, \sigma 
\left\{-2 \gamma_{\mu\nu}\frac{\delta}{\delta \gamma_{\mu\nu}}
+ \varphi \frac{\delta}{\delta \varphi}
+\Lambda \frac{\delta}{\delta \Lambda}
+ (d_n^0 -\hat{\gamma}_n)g_n^0\frac{\delta}{\delta g_n^0}
 \right\} S  =0
\; .
\eeq
If we rescale $g_n^0$ by $\Lambda^{d_n^0}$ such that $\rho_n$ are dimensionless,
\beq\label{eq:rho}
g_n^0=\Lambda^{d_n^0}\rho_n \qquad \text{with} \qquad
\Lambda\frac{\delta}{\delta \Lambda} \rho_n = {\gamma}_n(\Lambda, \rho_i), \qquad \gamma_n=(\hat{\gamma}_n - d_n^0)\,\rho_n\; ,
\eeq
eq.~\eqref{eq:rgen1} becomes
\beq\label{eq:rgenrho}
\int\sqrt{\gamma}d^d x\, \sigma 
\left\{-2 \gamma_{\mu\nu}\frac{\delta}{\delta \gamma_{\mu\nu}}
+ \varphi \frac{\delta}{\delta \varphi}
+\Lambda \frac{\delta}{\delta \Lambda}
 -{\gamma}_n\frac{\delta}{\delta \rho_n}
 \right\} S  =0
\; .
\eeq

Turning to case 2, we can relate the terms on the right hand side of eq.~\eqref{eq:rgen} to the usual $\beta$ functions of the $\mu$ dependent physical couplings $g_i$. To be specific, we can assume a momentum subtraction scheme where counter terms are defined by the condition that appropriate one-particle-irreducible Green functions evaluated at non-exceptional momenta $-p_i^2=\mu^2$ assume values of corresponding physical couplings $g_i$. We denote the canonical dimensions of $g_i$ by $d_i$, and apply naive dimensional analysis to $g_n^0(\Lambda,g_i,\mu)$:
\beq
\left\{\Lambda\frac{\delta}{\delta \Lambda} + \mu \frac{\partial}{\partial \mu}
+d_i g_i \frac{\delta}{\delta g_i}\right\} g_n^0=d_n^0\, g_n^0\; .
\eeq
Next we use the fact that the total derivative of $g_n^0(\Lambda,g_i,\mu)$ with respect to $\mu$ must vanish:
\beq
\mu \frac{\partial g_n^0}{\partial \mu} + \mu \frac{d g_i}{d \mu}\, \frac{\delta g_n^0}{\delta g_i} =0
\eeq
leading to
\beq\label{eq:mu}
\Lambda \frac{\delta g_n^0}{\delta \Lambda}=d_n^0\, g_n^0 +\left(\mu \frac{d g_i}{d \mu}-d_i\, g_i\right)
\frac{\delta g_n^0}{\delta g_i}\; .
\eeq
Thus the right hand side of eq.~\eqref{eq:rgen} becomes
\beq\label{eq:beta}
{\cal O}_n \beta_i\frac{\delta g_n^0}{\delta g_i}
\qquad \text{with}\qquad \beta_i=\mu \frac{d g_i}{d \mu}-d_i\, g_i\; .
\eeq
and eq.~\eqref{eq:rgen} can be written as
\beq\label{eq:rgen2}
\int\sqrt{\gamma}d^d x\, \sigma 
\left\{-2 \gamma_{\mu\nu}\frac{\delta}{\delta \gamma_{\mu\nu}}
+ \varphi \frac{\delta}{\delta \varphi}
+\Lambda \frac{\delta}{\delta \Lambda}
-\beta_i\frac{\delta}{\delta g_i}
 \right\} S  =0
\; .
\eeq
Here $S_{int}(g_n^0(\Lambda,g_i,\mu))$ is considered as $S_{int}(\Lambda,g_i,\mu)$, and the cutoff $\Lambda$ is also still present in the kinetic part $S_k$. We recall that eq.~\eqref{eq:rgen2} is valid before the limit $\Lambda \to \infty$ is taken.

\section{Gradient Flows from Weyl Consistency Conditions}

The aim of this section is to derive consequences of Weyl consistency conditions which follow from the local RGEs derived in Section~3. The scenarios case 1 and case 2 are considered separately, although the essential steps are the same.

\subsection{A Gradient Flow for Case 1}

In this case the action $S(\varphi,\Lambda,\rho_n)$ satisfies the local RGE eq.~\eqref{eq:rgenrho}.
We consider the vacuum partition function ${W^0}(\gamma_{\mu\nu},\Lambda,\rho_n)$ as functional of $\gamma_{\mu\nu},\Lambda,\rho_n$ given by
\beq\label{eq:W0}
e^{-{W^0}(\gamma_{\mu\nu},\Lambda,\rho_n)} = \frac{1}{\cal N}
\int {\cal D}\varphi\, e^{-S(\varphi,\gamma_{\mu\nu},\Lambda,\rho_n)}\; .
\eeq

Since $\varphi$ are dummy variables on the right hand side of eq.~\eqref{eq:W0}, eq.~\eqref{eq:rgenrho}
implies the local RGE
\beq\label{eq:localRGEW0}
\int\sqrt{\gamma}d^d x\, \sigma 
\left\{-2 \gamma_{\mu\nu}\frac{\delta}{\delta \gamma_{\mu\nu}}
+\Lambda\frac{\delta}{\delta \Lambda}
-{\gamma}_n\,\frac{\delta}{\delta \rho_n}
 \right\} W^0  =0
\eeq
with ${\gamma}_n$ defined in eq.~\eqref{eq:rho}. $W^0$ is invariant under general coordinate transformations and depends on $\Lambda(x)$, $\rho_n(x)$ and their derivatives. 

Since $\Lambda$ is local we express it in the form
\beq\label{eq:localLambda}
\Lambda(x) = \overline{\Lambda}\, e^{\lambda(x)}
\eeq
with $\overline{\Lambda}$=const.; the limit of an infinite cutoff would correspond to $\overline{\Lambda} \to \infty$, $\lambda(x)$ fixed, but is not considered here.
Under Weyl rescalings we have $\delta_\sigma \lambda = \sigma$, i.e.
\beq\label{eq:localLambda1}
\sigma \Lambda \frac{\delta}{\delta \Lambda} \equiv \sigma \frac{\delta}{\delta \lambda}\; .
\eeq

%Subsequently we focus on $d=4$ dimensions. Now $g_i$ are dimensionless, and masses $m_i$ will be treated separately (dropping cubic couplings for simplicity). 
%Since it will be convenient to work with dimensionless variables only, we expand $m_i(x)$ around constant values $\overline{m}_i$:
%\beq\label{eq:nu}
%m_i(x)=\overline{m}_i e^{\nu_i(x)}\; .
%\eeq
%The $\beta$ functions for $\nu_i(x)$ are defined such that $\beta_i\frac{\delta}{\delta g_i}$ in eq.~\eqref{eq:localRGEW0} becomes $\beta_{\nu_i} \frac{\delta}{\delta \nu_i}$:
%\beq
%\beta_{\nu_i}=\gamma_{m_i}-1\; , \qquad \gamma_{m_i}=\frac{\mu}{m_i}\frac{dm_i}{d\mu}\; .
%\eeq

Using $\lambda$ from eq.~\eqref{eq:localLambda} all local variables can be combined into dimensionless $\chi_i$ and corresponding $\beta$ functions $\tilde\beta_i$,
\beq
\chi_i=\{\lambda(x),\rho_n(x) \} \qquad \text{and}\qquad \tilde\beta_i = \{-1, \gamma_n(x)\}\; .
\eeq
Then eq.~\eqref{eq:localRGEW0} becomes
\beq\label{eq:localRGEW01}
\int\sqrt{\gamma}d^d x\, \sigma 
\left\{-2 \gamma_{\mu\nu}\frac{\delta}{\delta \gamma_{\mu\nu}}
-\tilde\beta_i\frac{\delta}{\delta \chi_i}
 \right\} W^0  =0\; .
\eeq

Subsequently we focus on $d=4$ dimensions, and we consider terms in $W^0$ involving four derivatives acting on the metric $\gamma_{\mu\nu}$ or $\chi_i$. A complete list of such structures respecting manifest coordinate invariance is given in [JO]. Five among these structures will play a role subsequently, given by
\bea
&&W^0=
\int\sqrt{\gamma}d^4x\,
\left\{b^0\,G+
\frac{1}{2}{\cal G}^0_{\chi_i\chi_j}\,G^{\mu\nu}\,\partial_\mu {\chi_i} \partial_\nu {\chi_j}+
{\cal E}^0_{\chi_i\chi_j}\,\partial^\mu R\, {\chi_i} \partial_\mu {\chi_j}+
\frac{1}{2}\,{\cal F}^0_{\chi_i\chi_j}\,R\,\partial_\mu {\chi_i} \partial^\mu {\chi_j}  \right.\nn\\
&&\left.
+\frac{1}{2}{\cal A}^0_{\chi_i\chi_j}\nabla^2 {\chi_i} \nabla^2 {\chi_j}
+ \dots
\right\}
\label{eq:Ws0}
\eea
where the coefficients $b^0, {\cal G}^0_{\chi_i\chi_j}, {\cal E}^0_{\chi_i\chi_j}, {\cal F}^0_{\chi_i\chi_j}$ and ${\cal A}^0_{\chi_i\chi_j}$ are functions of ${\Lambda}$ and ${\rho}_n$.
The upper index~$^0$ indicates that these coefficients are generally divergent for ${\overline{\Lambda}} \to \infty$.

$G$ is the Euler density and $G_{\mu\nu}$ the Einstein tensor:
\beq
G=R^{\alpha\beta\gamma\delta} R_{\alpha\beta\gamma\delta} -4R^{\alpha\beta}R_{\alpha\beta} +R^2,
\quad G_{\mu\nu}=R_{\mu\nu}-\frac{1}{2}\gamma_{\mu\nu} R\; .
\eeq
Under Weyl rescalings $\delta_\sigma \gamma_{\mu\nu}= -2\sigma \gamma_{\mu\nu}$, $G$ and $G_{\mu\nu}$ transform as
\beq
\delta_\sigma G = 4\sigma G-8G^{\mu\nu}\nabla_\mu \nabla_\nu \sigma\; , \quad
\delta_\sigma G_{\mu\nu}=2(\nabla_\mu \nabla_\nu \sigma -\gamma_{\mu\nu}\nabla^2 \sigma)\; .
\eeq

Applying the local RGE \eqref{eq:localRGEW01} to $W^0$ expanded as in \eqref{eq:Ws0} leads to terms $\sim \sigma$ and, after partial integrations and using $\nabla_\mu G^{\mu\nu}=0$, to terms $\sim \partial_\mu \sigma$ (see [JO]). One obtains an equation of the form
\beq\label{eq:W2}
\int\sqrt{\gamma}d^4x\, \left\{\sigma X + \partial_\mu\sigma Z^\mu\right\}=0
\eeq
where $X$ and $Z^\mu$ have to vanish separately. Alternatively, following \cite{Osborn:1991gm} one can introduce the operators
\beq
\Delta_\sigma^\gamma = \int \sqrt\gamma d^4x\left(-2\sigma \gamma_{\mu\nu}\frac{\delta}{\delta \gamma_{\mu\nu}}\right)\; ,\quad
\Delta_\sigma^\beta = \int \sqrt\gamma d^4x\left(\sigma \tilde{\beta}_i\frac{\delta}{\delta \chi_i}\right)
\eeq
and express eq.~\eqref{eq:localRGEW01} in the form
\beq
\left(\Delta_\sigma^\gamma-\Delta_\sigma^\beta\right)W^0=0\; .
\eeq
Since Weyl rescalings are abelian one must have
\beq\label{eq:abel}
\left[\Delta_\sigma^\gamma-\Delta_\sigma^\beta,\Delta_{\sigma'}^\gamma-\Delta_{\sigma'}^\beta\right]=0 \; .
\eeq
Acting with eq.~\eqref{eq:abel} on $W^0$ leads to the same conditions as we will obtain from eq.~\eqref{eq:W2}.

$X$ can be expanded in the same basis $\{ G,\; G^{\mu\nu}\partial_\mu \chi_i \partial_\nu \chi_j,\; \dots\}$ as $W^0$ in \eqref{eq:Ws0}. The required vanishing of each of the corresponding coefficients in $X$ requires
\bea
\tilde\beta_i \frac{\delta}{\delta \chi_i} b^0&=&0\nn \\
\tilde\beta_k \frac{\delta}{\delta \chi_k}
{\cal G}^0_{\chi_i\chi_j}
+{\cal G}^0_{\chi_k \chi_j}\frac{\delta \tilde\beta_k}{\delta \chi_i}
+{\cal G}^0_{\chi_k \chi_i}\frac{\delta \tilde\beta_k}{\delta \chi_j}
 &=& 0
\label{eq:simbeta}
\eea
and similar equations for ${\cal E}^0_{\chi_i\chi_j}$, ${\cal F}^0_{\chi_i\chi_j}$ and ${\cal A}^0_{\chi_i\chi_j}$. It is straightforward to separate the derivatives with respect to $\lambda$ in these equations which can then be interpreted as definitions of $\beta$ functions which describe the dependence of the bare coefficients $b^0$ and ${\cal G}^0_{ij}$ on the cutoff ${\Lambda}$:
\bea
&&\beta_b \equiv {\Lambda}\frac{\delta b^0}{\delta {\Lambda}} = \gamma_n \frac{\delta b^0}{\delta \rho_n} \nn \\
&&\beta_{{\cal G}_{nm}} \equiv {\Lambda}\frac{\delta {\cal G}^0_{\chi_n\chi_m}}{\delta {\Lambda}} =
\gamma_k \frac{\delta {\cal G}^0_{\chi_n\chi_m}}{\delta \rho_k}
+{\cal G}^0_{\chi_k \chi_m}\frac{\delta \gamma_k}{\delta \rho_n}
+{\cal G}^0_{\chi_k \chi_n}\frac{\delta \gamma_k}{\delta \rho_m}\; .
\label{eq:simbetak}
\eea

More interesting in the following are the consistency conditions which follow from the vanishing of the coefficients of the terms $G^{\mu\nu}\partial_\nu\chi_i$ in $Z^\mu$ in eq.~\eqref{eq:W2}. First, the action of $\int \sqrt{\gamma} d^4y(-2 \gamma_{\mu\nu}\frac{\delta}{\delta \gamma_{\mu\nu}})$ on $\sqrt{\gamma}b^0 G$ gives
\beq
\int\hspace*{-1mm}
\sqrt{\gamma} d^4y\left(-2\sigma \gamma_{\mu\nu}\frac{\delta}{\delta \gamma_{\mu\nu}}\right) \sqrt{\gamma} b^0 G=-8\sqrt{\gamma}b^0 G^{\mu\nu}\nabla_\mu \nabla_\nu \sigma
=8 \sqrt{\gamma} G^{\mu\nu} \partial_\mu \sigma \partial_\nu b^0
=8 \sqrt{\gamma} G^{\mu\nu} \partial_\mu \sigma  \partial_\nu \chi_i \frac{\delta b^0}{\delta \chi_i}
\eeq
where partial integration, $-2\sigma \gamma_{\mu\nu}\frac{\delta}{\delta \gamma_{\mu\nu}}\sqrt{\gamma}=-4\sigma \sqrt{\gamma}$
 and $\nabla_\mu G^{\mu\nu}=0$ have been used. 
Second, the action of $-\int\sqrt{\gamma} d^4y(\sigma\tilde\beta_i \frac{\delta}{\delta \chi_i})$ on the derivatives of $\chi_i$ in $\frac{1}{2}{\cal G}^0_{\chi_i\chi_j}\,G^{\mu\nu}\,\partial_\mu {\chi_i} \partial_\nu {\chi_j}$ gives (dropping uninteresting terms)
\beq
-\sqrt{\gamma} G^{\mu\nu} \partial_\mu \sigma  \partial_\nu \chi_i {\cal G}^0_{\chi_i\chi_j} \tilde\beta_j + \dots\; .
\eeq
These are the only terms proportional to $\sqrt{\gamma} G^{\mu\nu} \partial_\mu \sigma  \partial_\nu \chi_i$ after applying the local RGE \eqref{eq:localRGEW01} to $W^0$, hence the sum of the corresponding coefficients must vanish:
\beq\label{eq:gradflow}
8\frac{\delta b^0}{\delta \chi_i} -{\cal G}^0_{\chi_i\chi_j} \tilde\beta_j =0\; .
\eeq
This equation is of the form of an exact gradient flow for the $\beta$ functions $\tilde\beta_i \equiv \{-1, \gamma_n\}$, with
${\cal G}^0_{\chi_i\chi_j}$ a metric in the space $\chi_i\equiv\{\lambda,\rho_n\}$. Due to the introduction of dimensionless $\chi_i$, all components of ${\cal G}^0_{\chi_i\chi_j}$ are dimensionless. However, this form of the gradient flow makes sense only for a finite cutoff since both the metric ${\cal G}^0_{\chi_i\chi_j}$ and the potential $b^0$ will generally be divergent.

Let us consider the components $\chi_i=\rho_n$ of eq.~\eqref{eq:gradflow},
\beq\label{eq:gradflowg}
8\frac{\delta b^0}{\delta \rho_n} -{\cal G}^0_{\rho_n \rho_m} \gamma_m  +{\cal G}^0_{\rho_n\lambda}  =0
\eeq
where we can replace in the arguments of $b^0$ and ${\cal G}^0$ all couplings $\rho$ by constants $\bar{\rho}$.

In Appendix B (based on Appendix A) it is shown under which condition ${\cal G}^0_{\rho_n\lambda}(\bar{\rho})$ ($={\cal G}^0_{\lambda \rho_n}(\bar{\rho}$)) can be written as a gradient of a function ${\cal B}^0(\bar{\rho})$. If this condition is satisfied one finds an exact gradient flow for $\chi_i=\bar{\rho}_n$ of the form
\beq\label{eq:gradflowg1}
8\frac{\delta \hat{b}^0}{\delta \bar{\rho}_n}={\cal G}^0_{\rho_n \rho_m} \gamma_m\; , \qquad
\hat{b}^0={b}^0 +\frac{1}{8}{\cal B}^0(\bar{\rho})\; .
\eeq

We recall again that here we are considering the case of a quantum field theory with finite cutoff. The dependence of the couplings $\rho_n$ on the cutoff is given by anomalous dimensions, and the dependence of $b^0$ and ${\cal G}^0_{\rho_n \rho_m}$ on the cutoff is given by eqs.~\eqref{eq:simbetak}. In the gradient flow for the anomalous dimensions in eq.~\eqref{eq:gradflow} couplings to terms involving derivatives of $\lambda(x)$ like ${\cal G}^0_{g_i\lambda}(\bar{\rho})$ do play a role, but in eq.~\eqref{eq:gradflowg1} such couplings appear no longer. Still, calculations of diagrams with vertices involving $\lambda(x)$ are required in order to compute ${\cal B}^0(\bar{\rho})$ according to the rules given in Appendix B.

\subsection{Gradient Flows for Case 2}

Next we consider a renormalizable quantum field theory in $d=4$ dimensions where the renormalized action $S(\gamma_{\mu\nu},g_i,\Lambda,\mu)$ satisfies the local RGE given in eq.~\eqref{eq:rgen2}. $g_i$ denote marginal or relevant couplings, but again it will be useful to work with dimensionless fluctuating fields $\chi_i$ only. Let us assume marginal couplings and masses $m_i$ only, and expand $m_i(x)$ around constant values $\overline{m}_i$:
\beq\label{eq:nu}
m_i(x)=\overline{m}_i e^{\nu_i(x)}\; .
\eeq
The $\beta$ functions for $\nu_i(x)$ are defined such that $\beta_i\frac{\delta}{\delta g_i}$ in eq.~\eqref{eq:beta} becomes $(\gamma_{m_i}-1) \frac{\delta}{\delta \nu_i}$ with
\beq
 \gamma_{m_i}=\frac{\mu}{m_i}\frac{dm_i}{d\mu}\; .
\eeq
Again, using $\lambda$ from eq.~\eqref{eq:localLambda} all local variables can be combined into dimensionless $\chi_i$ and corresponding $\beta$ functions $\tilde\beta_i$,
\beq\label{bet1}
\chi_i=\{\lambda(x), \hat{g}_i(x)\}=\{\lambda(x), g_i(x),\nu_i(x)\} \qquad \text{and}\qquad 
\tilde\beta_i= \{-1, \hat{\beta}_i\} = \{-1, \beta_i,\gamma_{m_i}-1\}\; .
\eeq

The bare vacuum particion function is given by
\beq\label{rW0}
e^{-{W^0}(\gamma_{\mu\nu},\overline{\Lambda},\lambda,\hat{g}_i)} = \frac{1}{\cal N}
\int {\cal D}\varphi\, e^{-S(\varphi,\gamma_{\mu\nu},\Lambda,g_i^0(g_i,\Lambda,\mu),m_i^0(m_i,g_i,\Lambda,\mu))}\; .
\eeq
As before we consider terms of fourth order in derivatives acting on $\gamma_{\mu\nu}$, $\chi_i$, accordingly
$W^0$ can be expanded as in eq.~\eqref{eq:Ws0}. Eq.~\eqref{eq:rgen2} implies the local RGE for $W^0$
\beq\label{eq:localRGEW}
\int\sqrt{\gamma}d^d x\, \sigma 
\left\{-2 \gamma_{\mu\nu}\frac{\delta}{\delta \gamma_{\mu\nu}}
+\frac{\delta}{\delta \lambda}
-\beta_i\frac{\delta}{\delta g_i}
+(1-\gamma_{m_i})\frac{\delta}{\delta \nu_i}
 \right\} W^0  =0
\eeq
or, in terms of $\chi_i$ and $\tilde\beta_i$ defined in eq.~\eqref{bet1},
\beq\label{eq:localRGEW02}
\int\sqrt{\gamma}d^d x\, \sigma 
\left\{-2 \gamma_{\mu\nu}\frac{\delta}{\delta \gamma_{\mu\nu}}
-\tilde\beta_i\frac{\delta}{\delta \chi_i}
 \right\} W^0  =0\; .
\eeq

The counter terms implicit in $g_i^0(g_i,\Lambda,\mu)$ and $m_i^0(m_i,g_i,\Lambda,\mu)$ cancel subdivergences in the computation of $W^0$; however, superficial divergences for $\overline{\Lambda} \to \infty$ remain.
We have to add $W^{ct}$ which contains counter terms for the coefficients $b^0$, ${\cal G}^0_{\chi_i\chi_j}$, ${\cal E}^0_{\chi_i\chi_j}$, ${\cal F}^0_{\chi_i\chi_j}$ and ${\cal A}^0_{\chi_i\chi_j}$ in $W^0$ all of which are dimensionless. These counter terms $b^{ct}$, ${\cal G}^{ct}_{\chi_i\chi_j}$, ${\cal E}^{ct}_{\chi_i\chi_j}$, ${\cal F}^{ct}_{\chi_i\chi_j}$ and ${\cal A}^{ct}_{\chi_i\chi_j}$ depend on $\overline{\Lambda}$ such that the limit $\overline{\Lambda} \to \infty$ exists for $W=W^0+W^{ct}$. One obtains
\bea
&&e^{-{W}(\gamma_{\mu\nu},\lambda,\hat{g}_i)} = \left[\frac{1}{\cal N}
\int {\cal D}\varphi\, e^{-S(\varphi,\gamma_{\mu\nu},\Lambda,g_i^0(g_i,\Lambda,\mu),m_i^0(m_i,g_i,\Lambda,\mu))+W^{ct}(\gamma_{\mu\nu},\overline{\Lambda},\lambda,\hat{g}_i)}\right]_{\overline{\Lambda} \to \infty}\; ,\nn \\
&&W=
\int\sqrt{\gamma}d^4x\,
\left\{b\,G+
\frac{1}{2}{\cal G}_{\chi_i\chi_j}\,G^{\mu\nu}\,\partial_\mu {\chi_i} \partial_\nu {\chi_j}+
{\cal E}_{\chi_i\chi_j}\,\partial^\mu R\, {\chi_i} \partial_\mu {\chi_j}+
\frac{1}{2}\,{\cal F}_{\chi_i\chi_j}\,R\,\partial_\mu {\chi_i} \partial^\mu {\chi_j}  \right.\nn\\
&&\left.
+\frac{1}{2}{\cal A}_{\chi_i\chi_j}\nabla^2 {\chi_i} \nabla^2 {\chi_j}
+ \dots
\right\}\; .
\label{eq:Ws}
\eea

The argument $\lambda$ in $W(\gamma_{\mu\nu},\lambda,\hat{g}_i)$ indicates that $W$ still contains derivatives of $\lambda$ in terms like $\frac{1}{2}{\cal G}_{\lambda\lambda}\,G^{\mu\nu}\,\partial_\mu {\lambda}\, \partial_\nu {\lambda}$. Like all terms in $W^0$, ${\cal G}^0_{\lambda\lambda}$ has to be computed before the limit $\overline{\Lambda}\to \infty$ is taken. Being dimensionless, ${\cal G}^0_{\lambda\lambda}$ can contain powers of logarithms or remain finite for $\overline{\Lambda}\to \infty$; after adding ${\cal G}^{ct}_{\lambda\lambda}$,
${\cal G}_{\lambda\lambda}={\cal G}^0_{\lambda\lambda}+{\cal G}^{ct}_{\lambda\lambda}$ does not have to vanish for $\overline{\Lambda}\to \infty$. This remains true for all terms involving derivatives of $\lambda(x)$ in $W$ after renormalization.

$W^{ct}$ does not satisfy the local RGE \eqref{eq:localRGEW02}. Instead $W^{ct}$ satisfies a local RGE of the form
\beq\label{eq:localRGEWct}
\int\sqrt{\gamma}d^d x\, \sigma 
\left\{-2 \gamma_{\mu\nu}\frac{\delta}{\delta \gamma_{\mu\nu}}
-\tilde\beta_i\frac{\delta}{\delta \chi_i}
 \right\} W^{ct}  =-\int\sqrt{\gamma}d^d x\, \sigma {\cal R}_\beta
\eeq
with
\beq
{\cal R}_\beta=\beta^b G + \beta^{\cal G}_{\chi_i\chi_j} G^{\mu\nu} \partial\chi_i \partial \chi_j +\dots\; .
\eeq
(Using recursion relations as in the case of the standard renormalization theory it should be possible to show that the $\beta$ functions $\beta^b, \beta^{\cal G}_{\chi_i\chi_j}$ etc. are finite for $\overline{\Lambda}\to \infty$.) Accordingly $W$ satisfies the local RGE
\beq\label{eq:localRGEWW}
\int\sqrt{\gamma}d^d x\, \sigma 
\left\{-2 \gamma_{\mu\nu}\frac{\delta}{\delta \gamma_{\mu\nu}}
-\tilde\beta_i\frac{\delta}{\delta \chi_i}
 \right\} W  =\int\sqrt{\gamma}d^d x\, \sigma {\cal R}_\beta\; .
\eeq
Hence, instead of the consistency condition eq.~\eqref{eq:W2} one obtains
\beq\label{eq:cc}
\int\sqrt{\gamma}d^4x\, \left\{\sigma X + \partial_\mu\sigma Z^\mu\right\}=\int\sqrt{\gamma}d^d x\, \sigma {\cal R}_\beta
\eeq
where $X$ and $Z$ have to be computed as before but in terms of $b$ and ${\cal G}_{\chi_i\chi_j}$.
For terms $\sim \sigma G$ one obtains now (using $\frac{\delta b}{\delta \lambda}=0$)
\beq\label{eq:betab}
\beta^b=\hat\beta_i \frac{\delta b}{\delta \hat{g}_i}\; ,
\eeq
for terms $\sim \sigma G^{\mu\nu}\partial_\mu \hat{g}_i \partial_\nu \hat{g}_j$, $\sigma G^{\mu\nu}\partial_\mu \hat{g}_i \partial_\nu \lambda$ and $\sigma G^{\mu\nu}\partial_\mu \lambda \partial_\nu \lambda$ one gets using $\frac{\delta {\cal G}_{\chi_i\chi_j}}{\delta \lambda}=0$ and $\frac{\delta \hat\beta_i}{\delta \lambda}=0$
\bea
\beta^{\cal G}_{\hat{g}_i\hat{g}_j}&=&
\hat\beta_k \frac{\delta{\cal G}_{\hat{g}_i\hat{g}_j}}{\delta \hat{g}_k}
+{\cal G}_{\hat{g}_k \hat{g}_j}\frac{\delta \hat\beta_k}{\delta \hat{g}_i}
+{\cal G}_{\hat{g}_k \hat{g}_i}\frac{\delta \hat\beta_k}{\delta \hat{g}_j}\; ,\nn \\
\beta^{\cal G}_{\hat{g}_i\lambda}&=&
\hat\beta_k \frac{\delta{\cal G}_{\hat{g}_i\lambda}}{\delta \hat{g}_k}
+{\cal G}_{\hat{g}_k \lambda}\frac{\delta \hat\beta_k}{\delta \hat{g}_i}\; ,\nn \\
\beta^{\cal G}_{\lambda\lambda}&=&
\hat\beta_k \frac{\delta{\cal G}_{\lambda\lambda}}{\delta \hat{g}_k}\; .
\label{eq:simbetaW}
\eea

For terms proportional to $\sqrt{\gamma} G^{\mu\nu} \partial_\mu \sigma  \partial_\nu \chi_i$ in eq.~\eqref{eq:cc} one obtains no contribution from the right hand side. The terms with $\chi_i=\lambda, \hat{g}_i$ give, respectively,
\beq\label{eq:gradflowW}
8\frac{\delta b}{\delta \lambda}=0 ={\cal G}_{\lambda\chi_j} \tilde\beta_j = {\cal G}_{\lambda\hat{g}_j} \hat\beta_j - {\cal G}_{\lambda\lambda}\; ,\qquad
8\frac{\delta b}{\delta \hat{g}_i} ={\cal G}_{\hat{g}_i\chi_j} \tilde\beta_j = {\cal G}_{\hat{g}_i\hat{g}_j} \hat\beta_j - {\cal G}_{\hat{g}_i\lambda}\; .
\eeq
Again, if the condition discussed in Appendix~B under which ${\cal G}_{\hat{g}_i\lambda}(\hat{g})$ can be written as a gradient of a function ${\cal B}(\hat{g})$ is satisfied one finds an exact gradient flow of the form
\beq\label{eq:gradflowg3}
8\frac{\delta {\hat{b}}}{\delta \hat{g}_i}={\cal G}_{\hat{g}_i \hat{g}_j} \hat\beta_j\; , \qquad
\hat{b}={b} +\frac{1}{8}{\cal B}(\hat{g})\; .
\eeq

If one applies $\hat\beta_k\frac{\delta}{\delta \hat{g}_k}$ to each term in eq.~\eqref{eq:gradflowW} and uses eqs.~\eqref{eq:betab} and \eqref{eq:simbetaW} one obtains after some algebra
\beq\label{eq:gradflowbet}
8\frac{\delta \beta^b}{\delta \hat{g}_i} =\beta^{\cal G}_{\hat{g}_i g_j} \hat\beta_j  - \beta^{\cal G}_{\hat{g}_i\lambda}\; .
\eeq
This version of the Weyl consistency conditions is similar to eq.~(3.17a) in ref.~\cite{Jack:1990eb}.
Under the condition discussed in Appendix~B, $\beta^{\cal G}_{\hat{g}_i\lambda}$ can also be written as a gradient of a function $\beta^{\cal B}(\hat{g})=\frac{\delta}{\delta \hat{g}_k}(\hat\beta_k{\cal B}(\hat{g}))$ and one can find an exact gradient flow of the form
\beq\label{eq:gradflowg4}
8\frac{\delta {\hat{\beta}^b}}{\delta \hat{g}_i}=\beta^{\cal G}_{\hat{g}_i \hat{g}_j} \tilde\beta_j\; , \qquad
\hat{\beta}^b=\beta^b +\frac{1}{8}\beta^{\cal B}(\hat{g})\; .
\eeq

To sum up this section, Weyl consistency conditions implying gradient flows have been derived for the following scenarios:

1) For anomalous dimensions $\gamma_i$ for bare couplings $\rho_i$ associated to operators of any canonical dimensions, in terms of bare coefficients $b^0$ and ${\cal G}^0_{\chi_i\chi_j}$ in the form of eq.~\eqref{eq:gradflow} in the space $\chi_i=\{\lambda,\rho_i\}$. Under the condition discussed in Appendix~B, one obtains a gradient flow in the form of eq.~\eqref{eq:gradflowg1} in the space $\rho_i$ without $\lambda$.

2) For $\beta$ functions and anomalous dimensions for physical couplings and masses in renormalizable theories, approximate gradient flows in terms of renormalized coefficients $b$ and ${\cal G}_{\chi_i\chi_j}$ ($\chi_i=\{\lambda,\hat{g}_i\}$) and their $\beta$ functions in the form of eqs.~\eqref{eq:gradflowW} and \eqref{eq:gradflowbet}. Under the condition discussed in Appendix~B, exact gradient flows in the space $\hat{g}_i$ are of the form of eqs.~\eqref{eq:gradflowg3} and \eqref{eq:gradflowg4}.

\section{A Single Massive Scalar}

As a first application of the formalism (more will follow in a separate paper) we consider a single scalar $\varphi$ in $d=4$ dimensions with $g\varphi^4$ interaction and mass $m$. Given the Wilsonian cutoff $\Lambda$, quadratic UV divergences have to be cancelled by counter terms in $m_0^2$. In order to get a first idea about the formalism including mass terms we focus first on Weyl consistency conditions to lowest non-trivial order $\sim g^1$ for the bare couplings $b^0$ and ${\cal G}^0_{\chi_i\chi_j}$ (where $\chi_i = \lambda,\nu$) in the form of eq.~\eqref{eq:gradflow}.

For the kinetic part of the action we use eq.~\eqref{eq:kin} supplemented by a mass term,
\beq\label{eq:skm}
S_k=\frac{1}{2}\int\sqrt{\gamma}d^4x\,\left( \varphi(-\nabla^2+\frac{1}{6} R)\, F(D_\Lambda) \varphi + {m_0}^2\varphi^2\right)\, , \qquad
F(D_\Lambda)=e^{-D_\Lambda}\, ,
\eeq
and the interaction is
\beq
S_{int}=\int\sqrt{\gamma}d^4x\, \frac{g^0}{4!}{\varphi}^4
\eeq
where $g^0$ is the bare coupling. To ${\cal O}(g^1)$ eqs.~\eqref{eq:gradflow} read explicitely
\bea
8\frac{\delta b^0}{\delta \nu}
&=&\left(\gamma_m-1\right){\cal G}^0_{\nu\nu}  - {\cal G}^0_{\nu\lambda} \; ,\nn \\
8\frac{\delta b^0}{\delta \lambda}
&=&\left(\gamma_m-1\right){\cal G}^0_{\nu\lambda} - {\cal G}^0_{\lambda\lambda} \; .
\label{eq:cons1}
\eea

The calculation of ${\cal G}^0_{\chi_i\chi_j}$ proceeds via weak field expansions in $\gamma_{\mu\nu}=\eta_{\mu\nu}+h_{\mu\nu}$ to ${\cal O}(h_{\mu\nu})$ and in $\chi_i$ to ${\cal O}(\chi_i(x)^2)$, see Appendix~A, hence we need the vertices to the appropriate order given in Appendix~C.

The required counter terms in $m_0^2$ to ${\cal O}(g^1)$ are well known, but here we have to pay attention to the local nature of $m(x),\Lambda(x)$ in the form of eqs.~\eqref{eq:localLambda} and \eqref{eq:nu}. To this first order in $g$ we have to ${\cal O}(\chi_i(x)^2)$
\bea
m_0^2&=&m^2-\frac{g}{32\pi^2}\left(\Lambda^2-m^2\log\left(\Lambda^2/\mu^2\right)\right) \nn\\
&\simeq&\overline{m}^2(1+2\nu+2\nu^2) \nn\\
&&-\frac{g}{32\pi^2}\left(\overline{\Lambda}^2(1+2\lambda+2\lambda^2)
-\overline{m}^2(1+2\nu+2\nu^2)\log\left(\overline{\Lambda}^2/\mu^2\right)
-\overline{m}^2(2\lambda+4\nu\lambda)\right)
\label{eq:cts}
\eea
and
\beq
\gamma_m=\frac{g}{32\pi^2}\; .
\eeq

Whereas the vertices in $S_k$ involving $\lambda,\nu$ from $m_0^2$ in eq.~\eqref{eq:cts} are momentum independent, relatively complicated momentum dependent vertices follow from the term $\sqrt{\gamma}\varphi(-\nabla^2+\frac{1}{6} R)\, e^{-D_\Lambda} \varphi$ which has to be expanded to ${\cal O}(h_{\mu\nu}(x) \lambda(x)^2)$. These vertices are derived in Appendix~C. 

Diagrams contributing to ${\cal G}^0_{\chi_i\chi_j}$ to ${\cal O}(g^1)$ have the topology ``$\infty$'' with vertices involving the background fields $h_{\mu\nu}, \lambda, \nu$ attached to the loops. The free propagator $P(q)$ includes the UV cutoff,
\beq\label{eq:prop}
P(q)=1/(q^2 e^{q^2/\overline{\Lambda}^2}+m^2)\; .
\eeq
The loop integrals cannot be expressed in terms of standard functions; we will give the results in an expansion in $u \equiv \overline{m}^2/\overline{\Lambda}^2$.

For the free theory we find using the vertices in Appendix~C and the procedure described in Appendix~A
\beq
{\cal G}_{\nu\nu}^{0}=\frac{1}{16\pi^2} \frac{1}{45}+{\cal O}(u)
\label{eq:gnunu0}
\eeq 
whereas ${\cal G}_{\nu\lambda}^{0}$ is suppressed by $u$. The coefficients of terms suppressed by $u$ depend on the cutoff function $F(D_\Lambda)$ in $S_k$.
With $\beta_\nu=-1$ to lowest order it follows from eq.~\eqref{eq:cons1} 
\beq
\frac{\delta b^0}{\delta \nu} = -\frac{1}{16\pi^2} \frac{1}{360}+{\cal O}(u)\; .
\eeq
$b^0$ is divergent, and on dimensional grounds $b^0$ has to depend on $\Lambda/m$. Accordingly
\beq
b^0=\frac{1}{16\pi^2} \frac{1}{360}(\lambda-\nu)=\frac{1}{16\pi^2} \frac{1}{360} \left(\log\left(\Lambda/m\right) +\; \text{const.}\right)\; ,
\eeq
hence $b^0$ requires a counter term $b^{ct}$ such that
\beq
b=b^0+b^{ct}=\frac{1}{16\pi^2} \frac{1}{360} \left(\log\left(\mu/m\right) +\; \text{const.}\right)\; .
\eeq
It follows
\beq
\beta_b = \mu \frac{d b}{d \mu} = \frac{1}{16\pi^2} \frac{1}{360} 
\eeq
which coincides with the known value for a single scalar field.

To ${\cal O}(g^1)$ potential quadratic divergences are cancelled by counter terms. Somewhat astonishingly, finite contributions of ${\cal O}((u)^0)$ cancel as well. The subleading contributions $\Delta {\cal G}^0_{\chi_i\chi_j}$ of ${\cal O}(g^1)$ are
\bea
\Delta {\cal G}_{\nu\nu}^{0}&=&\frac{g}{(16\pi^2)^2} \frac{u}{45}\log\left(\frac{m^2}{\mu^2}\right)\; , \nn \\
\Delta {\cal G}_{\nu\lambda}^{0}&=&\frac{g}{(16\pi^2)^2} \frac{2u}{15}\log\left(\frac{m^2}{\mu^2}\right)\; , \nn \\
\Delta {\cal G}_{\lambda\lambda}^{0}&=&-\frac{g}{(16\pi^2)^2} \frac{13u}{45}\log\left(\frac{m^2}{\mu^2}\right)
\eea
and vanish for $\overline{\Lambda} \to \infty$, hence counter terms $\Delta {\cal G}^{ct}_{\chi_i\chi_j}$ are not required. (The logarithms $\log\left({m^2}/{\mu^2}\right)$ originate from the counter terms in $m_0^2$). The last term has been deduced from eqs.~\eqref{eq:cons1} using again, since $b^0$ is dimensionless, ${\delta b^0}/{\delta \nu} =- {\delta b^0}/{\delta \lambda}$.
Given the leading contribution to ${\cal G}^0_{\nu\nu}$ in eq.~\eqref{eq:gnunu0} the leading metric in the subspace $\nu, \lambda$ is always positive.

The subleading contributions show no sign for a diagonal metric ${\cal G}^0_{\chi_i\chi_j}$ in the subspace $\{\chi_i=m,$ $\Lambda\}$, but these are scheme dependent, i.e. dependent on the form of the cutoff function $F(D_\Lambda)$ in $S_k$.  Since the Weyl consistency conditions are covariant under redefinitions in the space $\chi_i$ (provided that $\beta$ functions are appropriately redefined), diagonalization of the metric is straightforward. The cancellation of finite contributions of ${\cal O}((u)^0)$ to $\Delta {\cal G}^0_{\chi_i\chi_j}$ may be an artefact of the low order in the coupling considered here.

\section{Conclusions and Outlook}
 
In the present article we have derived Weyl consistency conditions in the framework of a Wilsonian cutoff implemented in the kinetic term for scalar fields. The crucial tool is a local cutoff $\Lambda(x)$ which transforms under Weyl rescalings and allows for Weyl invariant kinetic terms. This formalism provides an alternative to Weyl consistency conditions derived using dimensional regularization by~[JO]. 

It allows to define the running of bare couplings with the UV cutoff via anomalous dimensions, a concept which makes no sense in dimensional regularization. Also it allows to treat couplings of any canonical dimension on the same footing, see the case~1 above. On the other hand the local field $\lambda(x)$ describing the $x$-dependence of the cutoff contributes {\it a priori} to the Weyl consistency conditions in the form of an additional local coupling. In this enlarged space of local fields and corresponding anomalous dimensions Weyl consistency conditions assume the form of an exact gradient flow in eq.~\eqref{eq:gradflow} in Subsection~4.1. (For attempts to generalize the AdS/CFT correspondence towards general QFTs (for which an exact gradient flow is essential) a local cutoff could be related to a component in the $(d+1)$ dimensional metric corresponding to the extra dimension.)

Within the reduced space corresponding to couplings $\hat{g}$ only, an extra term involving ${\cal G}_{g_i\lambda}$ is still present in the Weyl consistency conditions which spoils the gradient flow property unless this term itself can be expressed as a gradient. This extra term depends on contributions from diagrams $\sim \partial_\mu \lambda(x)$ to the vacuum partition function, and we discuss in Appendix~B under which condition ${\cal G}_{g_i\lambda}$ can be expressed as a gradient. It is easy to see that this condition is satisfied in lowest nontrivial orders in perturbation theory, but general criteria for its satisfaction remain to be investigated.

In Subsection~4.2 we applied the formalism to renormalizable theories for which the limit $\overline{\Lambda} \to \infty$ can be taken. Still, contributions $\sim \partial_\mu \lambda(x)$ to the vacuum partition function remain and, as before, corresponding couplings remain present in the Weyl consistency conditions. Again, these become a gradient flow only if the condition discussed in Appendix~B is satisfied. Finally we expressed the Weyl consistency conditions in terms of $\beta$ functions for the coefficients in the vacuum partition function; in this form they are very similar to the ones obtained by [JO] except that marginal and relevant couplings appear together.

We have also introduced and verified a method to compute the ``metric'' ${\cal G}_{\chi_i\chi_j}$ in momentum space in a weak field expansion. The explicit calculations are straightforward but involved due to the cutoff propagators and the vertices originating from the local cutoff in the kinetic term. We have computed ${\cal G}_{\chi_i\chi_j}$ to lowest non-trivial order in the coupling of a $\varphi^4$ theory; further studies of higher orders, of other models and of other cutoff functions leading to potential simplifications are desirable in the future.

A way to simplify higher loop calculations would be to add the cutoff function \eqref{eq:FDe} also to the mass term. This would generate additional vertices involving $\lambda(x)$, but the free cutoff propagator \eqref{eq:prop} simplifies to
\beq\label{eq:props}
P(q)=e^{-q^2/\overline{\Lambda}^2}/(q^2 +m^2) = \int_{1/\overline{\Lambda}^2}^\infty d\alpha\, e^{-\alpha q^2}\; .
\eeq
Such a Schwinger-like representation allows to execute multi-loop momentum integrals, and one is left with truncated (overlapping) $\alpha$ integrals.

\newpage

\appendix

\section*{Appendix}

%\subsection*{A How to compute ${\cal G}_{\chi_i\chi_j}$}
\section{How to compute ${\cal G}_{\chi_i\chi_j}$}

The aim of this section is to describe how ${\cal G}_{\chi_i\chi_j}$, the coefficients of $G^{\mu\nu}\,\partial_\mu {\chi_i} \partial_\nu {\chi_j}$ in the vacuum partition function $W$, can be computed in momentum space. (The following steps hold both for ${\cal G}^0_{\chi_i\chi_j}$ in terms of $W^0$, and ${\cal G}_{\chi_i\chi_j}$ in terms of $W$.)
We assume that $W$ is computed diagrammatically as function of background fields $f_i(p_i)$ including the local couplings and the local cutoff, and the metric $\gamma_{\mu\nu}(q)$.

The first step is an expansion of the metric around flat space, $\gamma_{\mu\nu}=\eta_{\mu\nu}+h_{\mu\nu}$, and to consider the terms linear in $h_{\mu\nu}$:
\beq
W=\int\frac{d^4 q}{(2\pi)^4} h_{\mu\nu}(q)\, T^{\mu\nu}(-q,f_i)+\dots
\eeq
Next, expand $f_i(p_i)$ around constants in space-time, $f_i(p_i)=\bar{f}_i(2\pi)^4\delta^4(p_i)+\chi_i(p_i)$, and keep only terms of second order in $\chi_i(p_i)$ which correspond to the (Fourier transforms of) the local fields $\chi_i$ in section~4. Denoting the momenta of two fluctuations $\chi_i(p_i)$ by $p$ and $q-p$ using $\sum_i p_i=q$, $T^{\mu\nu}(-q)$ is of the form
\beq
T^{\mu\nu}(-q,f_i)=\int\frac{d^4 p}{(2\pi)^4}\, \chi_i(p)\chi_j(q-p)\, t^{\mu\nu}_{ij}(p,q)\; .
\eeq

We are interested in the terms quartic in derivatives, i.e. quartic in the momenta $p,q$ in $t^{\mu\nu}_{ij}(p,q)$. These terms can be decomposed into
\beq\label{eq:tk}
 t^{\mu\nu}_{ij}(p,q)=K_{1ij}p^\mu p^\nu +\frac{1}{2}K_{2ij}(p^\mu q^\nu+q^\mu p^\nu)
+K_{3ij}q^\mu q^\nu +K_{4ij}\eta^{\mu\nu}
\eeq
where $K_{1ij},K_{2ij},K_{3ij}$ are polynomials of first order in the Lorentz invariants $p^2$, $pq$, $q^2$, and $K_{4ij}$ is of second order in the same Lorentz invariants.
Next we have to expand the terms in $W^0$ or $W$ in eqs.~\eqref{eq:Ws0}/\eqref{eq:Ws} to the same order in $h_{\mu\nu}$ and $p,q$. The first term $\sim G$ drops out since of higher order in $h_{\mu\nu}$, and further terms not written explicitely in $W^0$/$W$ are of higher order in $\chi_i(p_i)$.  
The terms quartic in the momenta $p_i$ proportional to $h^{\mu\nu}(q)\chi_i(p)\chi_j(q-p)$ from the remaining four terms in eq.~\eqref{eq:Ws} are of the following form, once expressed in momentum space:
\begin{align}
G^{\rho\sigma}\,\partial_\rho {\chi_i} \partial_\sigma\, {\chi_j}:%& 
\frac{1}{2}&\left(q^2 p_{\mu}p_{\nu}-pq (p_{\mu}q_{\nu}+ q_{\mu}p_{\nu})+p^2 q_{\mu}q_{\nu}+\eta_{\mu\nu}((pq)^2-p^2q^2)\right)
\nn\\
\partial^\rho R\, {\chi_i} \partial_\rho {\chi_j}:%&
\frac{1}{2}&(-q_\mu q_\nu q^2+\eta_{\mu\nu} q^2 q^2)
\nn\\
R\,\partial_\rho {\chi_i} \partial^\rho {\chi_j}:%&
-&q_\mu q_\nu (p^2+pq)+\eta_{\mu\nu} q^2(p^2+pq)
\nn\\
\sqrt{\gamma}\nabla^2 {\chi_i} \nabla^2 {\chi_j}:%&
-&(p_\mu p_\nu+p_\mu q_\nu)(2p^2+2pq+q^2) +\eta_{\mu\nu}(\frac{1}{2}p^2 p^2+p^2 pq+pq^2+\frac{1}{2}pq q^2)
\end{align}
It follows that ${\cal G}_{\chi_i\chi_j},\, {\cal E}_{\chi_i\chi_j},\, {\cal F}_{\chi_i\chi_j},\, {\cal A}_{\chi_i\chi_j}$ can be obtained from $t^{\mu\nu}_{ij}(p,q,\bar\chi_k)$, decomposed as in eq.~\eqref{eq:tk}, as
\bea
{\cal G}_{\chi_i\chi_j}&=&2\frac{d}{dq^2}(K_{1ij}-K_{2ij})\nn \\
{\cal A}_{\chi_i\chi_j}&=&-\frac{d}{dq^2}K_{2ij}\nn \\
{\cal F}_{\chi_i\chi_j}&=&-\frac{d}{d(pq)}K_{3ij}\nn \\
{\cal E}_{\chi_i\chi_j}&=&\left(\frac{d}{dq^2}\right)^2K_{4ij}\; .
\label{eq:GAFE}
\eea
For our purposes the first relation for ${\cal G}_{\chi_i\chi_j}$ is all we need.

We note that this approach in momentum space is general, independent from the UV regularization used to compute the three point functions $\left<h_{\mu\nu}(q) \chi_i(p) \chi_j(q-p)\right>$. We have verified it for a scalar theory with $g(x) \varphi^4$ interaction in $d=4-\varepsilon$ dimensions, where the three point functions $\left<h_{\mu\nu}(q) g(p) g(q-p)\right>$ are calculable to three loop order. We found ${\cal G}_{g g}$
%=\frac{1}{(16\pi)^3 432 \varepsilon}$ 
in agreement with eq.~(6.30) in \cite{Jack:1990eb}.% (actually up to a sign).

\section{${\cal G}_{\lambda g_i}$ as a Gradient }

The aim of this section is to show under which condition ${\cal G}_{\lambda g_i}={\cal G}_{g_i \lambda}$ can be written as a gradient. The reasoning below applies to ${\cal G}^0_{\lambda \rho_n}$ in subsection 4.1, to ${\cal G}^0_{\lambda g_i}$, to the counterterm ${\cal G}^{ct}_{\lambda g_i}$ and hence to the renormalized ${\cal G}_{\lambda g_i}$ in subsection 4.2. For simplicity we use the notation ${\cal G}_{\lambda g_i}$ for all cases.

To start with, in momentum space $W$ depends on $\chi_i(p_i)\equiv \left\{\Lambda(p_\lambda), g_i(p_i)\right\}$. We assume $N$ distinct couplings $g_i$ ($i=1\dots N$). The momentum dependent fields have to be expanded around constants in space-time,
\beq\label{eq:exp}
\Lambda(p_\lambda)=\overline{\Lambda}(2\pi)^4\delta^4(p_\lambda)+\widetilde{\lambda}(p_\lambda)\, ,\qquad
g_i(p_i)=\bar{g}_i(2\pi)^4\delta^4(p_i)+\tilde{g}_i(p_i)\, .
\eeq 
Following Appendix~A, $W$ contains ${\cal G}_{\lambda g_i}$ in terms of the form
\beq\label{eq:B2}
W=\sum_{i=1}^N \int\frac{d^4 q}{(2\pi)^4} \frac{d^4 p_\lambda}{(2\pi)^4} \frac{d^4 p_i}{(2\pi)^4} \delta^4(q+p_\lambda+p_i)\,
h_{\mu\nu}(q)\, \widetilde{\lambda}(p_\lambda)\, \tilde{g}_i(p_i)\, q^2\, p^\mu_\lambda\, p_i^\nu\, {\cal G}_{\lambda g_i}(\bar{g}) + \dots\; .
\eeq

Accordingly ${\cal G}_{\lambda g_i}$ can be computed in terms of $W$ as follows: First, expand $W$ to linear order in $h_{\mu\nu}(q)$, $\widetilde{\lambda}(p_\lambda)$ and to ${\cal O}(q^2)$ and ${\cal O}(p^\mu_\lambda)$ in the corresponding momenta: 
\beq\label{eq:B3}
W=\int\frac{d^4 q}{(2\pi)^4} \frac{d^4 p_\lambda}{(2\pi)^4}\,
 h_{\mu\nu}(q)\, \widetilde{\lambda}(p_\lambda)\, q^2\, p^{\mu}_\lambda\, T^{\nu}(-q,g_i) + \dots\; .
\eeq
Here the local couplings $g_i$ in $T^{\nu}(-q,g_i)$ are not yet expanded around constants in space-time as in eq.~\eqref{eq:exp}.
We assume that $T^{\nu}(-q,g_i)$ can be written as a series in $n_i$ powers of $g_i(p_{i,j_i})$, $j_i=1\dots n_i$:
\beq\label{eq:tnu}
T^{\nu}(-q,g_i)=\int{\cal D}p\, \sum_{n_1 \dots n_N}\,\sum_k\, p_k^\nu\,  \delta^4(q+p+p_k)\, t_{n_1\dots n_N,k}\, \prod_{i=1}^N g_i^{n_i}(p_{i,j_i}) \; .
\eeq
In eq.~\eqref{eq:tnu}, $\int{\cal D}p$ denotes the integrals over all $n_1\times $\dots$ \times n_N$ momenta, the arguments of the couplings $g_i^{n_i}(p_{i,j_i})\equiv g_i(p_{i,1})\times $\dots$ \times g_i(p_{i,n_i})$.
The sum over $k$ runs over all possible choices for $p_k$ among these momenta. In general, the coefficients $t_{n_1\dots n_N,k}$ are different for momenta $p_k$ corresponding to different couplings $g_i$. The condition for ${\cal G}_{\lambda g_i}$ as a gradient is the following:

{\bf After summation of all diagrams which contribute to $t_{n_1\dots n_N,k}$ to a given order, $t_{n_1\dots n_N,k}$ are symmetric under the exchange of momenta $p_k$ originating from different vertices corresponding to the same kind of coupling~$g_i$.} 

This condition is satisfied if, before the expansion of $T^{\nu}(-q,g_i)$ to ${\cal O}(p_k^\nu)$ in eq.~\eqref{eq:tnu}, $T^{\nu}(-q,g_i)$ is symmetric under the exchange of momenta $p_k$ originating from different vertices corresponding to the same kind of coupling~$g_i$ which holds
at least in low orders in perturbation theory, but counter examples may exist.

Next one has to expand all couplings around constants $\bar{g}_i$ in space-time.
Only the terms linear in the fluctuation $\tilde{g}_i(p_i)$ contribute to eq.~\eqref{eq:B2}. 
Inserting the expansion \eqref{eq:exp} for $g_i(p_i)$ into eq.~\eqref{eq:tnu}, and under the above condition, one obtains 
$n_1$ contributions $\sim \tilde{g}_1(p_1,j_1)$ from $g_1^{n_{1}}(p_{1,j_1})$ with $j_1=1\dots n_1$, and
$n_2$ contributions $\sim \tilde{g}_2(p_2,j_2)$ from $g_2^{n_{2}}(p_{2,j_2})$, $j_2=1\dots n_2$ etc.. 

Consider the contributions involving $\tilde{g}_1$. Under the above condition,
all $n_1$ contributions $\sim \tilde{g}_1(p_{1,j_1}),\; j_1=1\dots n_1,$ have the same coefficients $t_{n_1\dots n_N,k}$. 
The momenta $p_k^\nu$ in eq.~\eqref{eq:tnu} correspond to $p_{1,j_1}$ since all other momenta vanish, and all momenta $p_{1,j_1}$ can be denoted by $p_{1}$. 

The same features hold for the $n_2$ contributions $\sim \tilde{g}_2(p_2,j_2)$ etc.. Finally $T^{\nu}(-q,g_i)$ becomes
\begin{align}
T^{\nu}(-q,g_i)&=\sum_{i=1}^N\, \int \frac{d^4 p_i}{(2\pi)^4}\, p_i^\nu\,  \delta^4(q+p+p_i)\, \tilde{g}_i(p_i)\,\sum_{n_1 \dots n_N} t_{n_1\dots n_N,i}\,   n_i\, \bar{g}_i^{n_i -1} \prod_{j=1, j \neq i}^N \bar{g}_j^{n_j} \nn \\
&=  \sum_{i=1}^N\, \int \frac{d^4 p_i}{(2\pi)^4}\, p_i^\nu\,  \delta^4(q+p+p_i)\,\tilde{g}_i(p_i) \, \sum_{n_1 \dots n_N} t_{n_1\dots n_N,i}\,  \frac{d}{d\bar{g}_i} \,\prod_{j=1}^N \bar{g}_j^{n_j}\nn \\
&=  \sum_{i=1}^N\, \int \frac{d^4 p_i}{(2\pi)^4}\, p_i^\nu\,  \delta^4(q+p+p_i)\,\tilde{g}_i(p_i) \,
\frac{d}{d\bar{g}_i} {\cal B}(\bar{g})
\label{eq:tnu1}
\end{align}
where $t_{n_1\dots n_N,i}$ denote the coefficients corresponding to couplings $g_i$, and
\beq
{\cal B}(\bar{g})=\sum_{n_1 \dots n_N} t_{n_1\dots n_N,i}\,   \,\prod_{j=1}^N \bar{g}_j^{n_j}\; .
\eeq

It remains to replace eq.~\eqref{eq:tnu1} for $T^{\nu}(-q,g_i)$ into eq.~\eqref{eq:B3} which becomes
\beq\label{eq:wf}
W= \sum_{i=1}^N\,
\int\frac{d^4 q}{(2\pi)^4} \frac{d^4 p_\lambda}{(2\pi)^4}\, \frac{d^4 p_i}{(2\pi)^4}\,
\delta^4(q+p+p_i)\, h_{\mu\nu}(q)\, \widetilde{\lambda}(p_\lambda)\, \tilde{g}_i(p_i) \, q^2\, p^{\mu}_\lambda\,  p_i^\nu \frac{d}{d\bar{g}_i} \, {\cal B}(\bar{g}) + \dots\; .
\eeq
Comparing eq.~\eqref{eq:wf} to eq.~\eqref{eq:B2} one obtains
\beq\label{eq:B8}
{\cal G}_{\lambda g_i}(\bar{g})= \frac{d}{d\bar{g}_i}{\cal B}(\bar{g})\; ,
\eeq
i.e. ${\cal G}_{\lambda g_i}(\bar{g})$ is the gradient of a function ${\cal B}(\bar{g})$ defined by
\beq\label{eq:calb}
W= \sum_{i=1}^N\,
\int\frac{d^4 q}{(2\pi)^4} \frac{d^4 p_\lambda}{(2\pi)^4}\, \frac{d^4 p_i}{(2\pi)^4}\,
\delta^4(q+p+p_i)\, h_{\mu\nu}(q)\, \widetilde{\lambda}(p_\lambda)\, \tilde{g}_i(p_i) \, q^2\, p^{\mu}_\lambda\,  p_i^\nu \, {\cal B}(\bar{g}) + \dots\; .
\eeq
The origin of this property is the expansion of $g_i(p_i)$ around constants $\bar{g_i}$, and the expansion of $W$ to first order in $\tilde{g}_i(p_i)$ which converts $W$ as function of $g_i(p_i)$ into derivatives with respect to $\bar{g_i}$. 

The above reasoning applies also to ${\cal G}^0_{\lambda g_i}(\overline{\Lambda})$ and ${\cal G}^{ct}_{\lambda g_i}(\overline{\Lambda})$. After renormalization eq.~\eqref{eq:B8} holds for
${\cal G}_{\lambda g_i}(\bar{g})$=$\left[ {\cal G}^0_{\lambda g_i}(\bar{g},\overline{\Lambda}) - {\cal G}^{ct}_{\lambda g_i}(\bar{g},\overline{\Lambda})\right]_{\overline{\Lambda}\to \infty}$ and 
%\newline
 ${\cal B}(\bar{g})$=$\left[{\cal B}^0(\bar{g},\overline{\Lambda})-{\cal B}^{ct}(\bar{g},\overline{\Lambda}) \right]_{\overline{\Lambda}\to \infty} $.

%\newpage

\section{Vertices from $S_k$}

For the calculation of ${\cal G}_{\chi_i\chi_j}$, $\chi_i/\chi_j=\lambda$, one needs the vertices from $S_k$ quadratic in $\varphi$, up to linear order in $h_{\mu\nu}$ and up to quadratic order in $\lambda$ in momentum space, for our choice $F(D_\Lambda) = e^{-D_\Lambda}$ (recall $D_\Lambda = \Lambda^{-2}(\nabla^2-\frac{1}{6} R)$). It is convenient to split the terms in $S_k$ as
\beq\label{eq:C1}
\sqrt{\gamma} \varphi_l \left(-\nabla^2+\frac{1}{6} R\right) e^{-D_\Lambda}\varphi_r \equiv 
\varphi_l\,(-\square e^{-u\square})\,\varphi_r
+\varphi_l\left( V_l e^{-D_\Lambda}+\overleftarrow{\square} V_r\right) \varphi_r
\eeq
where
\beq
u=1/\overline{\Lambda}^2\,  ,\qquad
V_l = \sqrt{\gamma}\left(-\overleftarrow{\nabla}^2+\frac{1}{6} R\right)+\overleftarrow{\square}\, ,\qquad
V_r=e^{-u\square}-e^{-D_\Lambda}\; .
\eeq
The first term $(-\square e^{-u\square})$ in eq.~\eqref{eq:C1} defines the free propagator including the UV cutoff, $V_l$, $V_r$ and $e^{-D_\Lambda}$ contain vertices, the derivatives in $V_l$ act to the left on $\varphi_l$ and the derivatives in $V_r$ and in $e^{-D_\Lambda}$ on $\varphi_r$ to the right.

To linear order in $h_{\mu\nu}$ we have ($h\equiv h_\mu^{\ \ \mu}$)
\beq
V_l= -\frac{1}{2}\overleftarrow{\square}h-\frac{1}{2}\overleftarrow{\partial}_\mu(\partial_\mu h)
+\frac{1}{2}\overleftarrow{\partial}_\mu(\partial_\nu h^{\mu\nu})
+\overleftarrow{\partial}_\mu\overleftarrow{\partial}_\nu h^{\mu\nu}
+\frac{1}{6}\left((\partial_\mu\partial_\nu h^{\mu\nu})-(\square h)\right)\; .
\eeq
Subsequently, in momentum space, the derivatives acting to the left have to be replaced by $i$ times the momentum $l_1$ of $\varphi_l(l_1)$ and the derivatives acting on $h_{\mu\nu}$ by $i$ times the momentum $p_h$ of $h_{\mu\nu}(p_h)$. 

$e^{-D_\Lambda}$ can be expanded to linear order in $h^{\mu\nu}$ in the exponent,
\beq
e^{-D_\Lambda}=e^{-\Lambda^{-2} B- \Lambda^{-2} \square}
\eeq
with
\beq
B=\frac{1}{2}(\partial_\mu h){\partial}_\mu
-(\partial_\mu h^{\mu\nu}){\partial}_\nu
- h^{\mu\nu} {\partial}_\mu{\partial}_\nu
+\frac{1}{6}\left((\square h)-(\partial_\mu\partial_\nu h^{\mu\nu})\right)\; .
\eeq
Subsequently we expand $e^{-\Lambda^{-2} B- \Lambda^{-2} \square}$ to linear order in $h^{\mu\nu}$ using
\beq\label{eq:CA}
e^{C+A}\vert_{{\cal O}(C)}=\left(C+\frac{1}{2}[A,C]+\frac{1}{6}[A,[A,C]]+\frac{1}{24}[A,[A,[A,C]]]+\dots\right) e^A
\eeq
for $A=- \Lambda^{-2} \square$, $C=-\Lambda^{-2} B$. The commutators generate derivatives acting on $\Lambda^{-2}(x)$ resp. $\lambda(x)$ in $A$. From the projection of diagrams on ${\cal G}_{\chi_i\chi_j}$ from Appendix~A we know that we need at most 3 derivatives acting on $\lambda(x)$, therefore we can truncate the expansion in commutators at this order. Explicitely one obtains
%\bea
\begin{align}
e^{-D_\Lambda}|_{{\cal O}(B)}=&\, \Big\{-\Lambda^{-2}B+\frac{1}{2}\left(\Lambda^{-2}\square\Lambda^{-2}B-\Lambda^{-2}B\Lambda^{-2}\square\right) \nn\\
&-\frac{1}{6}\left(\Lambda^{-2}\square(\Lambda^{-2}\square\Lambda^{-2}B-\Lambda^{-2}B\Lambda^{-2}\square) -(\Lambda^{-2}\square\Lambda^{-2}B-\Lambda^{-2}B\Lambda^{-2}\square)\Lambda^{-2}\square\right) \nn\\
& +\frac{1}{24}\Big(
\Lambda^{-2}\square(\Lambda^{-2}\square(\Lambda^{-2}\square\Lambda^{-2}B-\Lambda^{-2}B\Lambda^{-2}\square) -(\Lambda^{-2}\square\Lambda^{-2}B-\Lambda^{-2}B\Lambda^{-2}\square)\Lambda^{-2}\square)
 \nn\\
&
 -(\Lambda^{-2}\square(\Lambda^{-2}\square\Lambda^{-2}B-\Lambda^{-2}B\Lambda^{-2}\square) -(\Lambda^{-2}\square\Lambda^{-2}B-\Lambda^{-2}B\Lambda^{-2}\square)\Lambda^{-2}\square)\Lambda^{-2}\square
\Big)\Big\} \nn\\
&\times e^{- \Lambda^{-2} \square} 
\label{eq:vr}
%\eea
\end{align}
where, from $\lambda(x)= \log(\Lambda/\overline{\Lambda})$ and to ${\cal O}(\lambda^2)$, $\Lambda^{-2}\simeq u(1-2\lambda(x)+2\lambda^2(x))$. The remaining exponential function in eq.~\eqref{eq:vr} remains to be expanded to ${\cal O}(\lambda^2)$. Again we write it in the form 
\beq\label{eq:aprime}
e^{C'+A'}\, ,\qquad A'=-u\square\; ,\qquad C'=-u(-2\lambda(x)+2\lambda^2(x))\; .
\eeq
Terms of ${\cal O}(C')$ can be obtained from eq.~\eqref{eq:CA}, but we need also terms of ${\cal O}(C'^2)$. These can be found in ref.~\cite{Kimura:2017xxz}. We introduce
\beq
{\cal B}_n = ({\cal L_{A'}})^{n-1} C'\qquad \text{where}\qquad {\cal L_{A'}}{\cal O} = [A',{\cal O}]\; .
\eeq
Then the terms of ${\cal O}(C'^2)$ in the expansion of $e^{C'+A'}$ read
\begin{align}
e^{C'+A'}|_{{\cal O}(C'^2)}=&\; \Bigg\{
{\cal B}_1\left(\frac{1}{2}{\cal B}_1+\frac{2}{3}{\cal B}_2+\frac{3}{4}{\cal B}_3\right)
+{\cal B}_2\left(\frac{1}{3}{\cal B}_1+\frac{1}{2}{\cal B}_2+\frac{3}{5}{\cal B}_3\right)\nn \\
&+{\cal B}_3\left(\frac{1}{4}{\cal B}_1+\frac{2}{5}{\cal B}_2+\frac{1}{2}{\cal B}_3\right)
\Bigg\}e^{A'}
\end{align}
(note that the order of different ${\cal B}_n$ matters). Replacing $A'$ and $C'$ from eq.~\eqref{eq:aprime} one obtains, similarly to eq.~\eqref{eq:vr}, a series of Laplacians $\square$ acting on $(-2\lambda(x)+2\lambda(x)^2)$ instead of $B$. 

It remains to expand all contributions to $e^{-D_\Lambda}$ resp. $V_r$ to ${\cal O}(\lambda^2(x))$, keeping track of the positions of $\square$ and $B$. In momentum space, all derivatives have to be replaced by $i$ times the sum of the momenta of $h_{\mu\nu}(p_h)$ in $B$, and the momenta in $\varphi_r(l_2)$ and $\lambda(p_1)$ (or $\lambda(p_2)$) for terms to the right of each derivative. Finally the full vertices $V$ are obtained by adding $\left( V_l e^{-D_\Lambda}+\square V_r\right)=V$ and dropping unnecessary terms of ${\cal O}(h^2)$ and ${\cal O}(\lambda^3)$.

Let us decompose $V$ into
\beq
V=\lambda(p_1)V^\lambda(p_1)+\lambda(p_1)\lambda(p_2)V^{\lambda\lambda}(p_1,p_2)+h_{\mu\nu}(p_h)V^h_{\mu\nu}(p_h)
+h_{\mu\nu}(p_h)\lambda(p_1)V^{h\lambda}_{\mu\nu}(p_h,p_1)
\eeq
where it is understood that each component depends in addition on the momenta $l_1,l_2$ of $\varphi_l(l_1)$, $\varphi_r(l_2)$ and $V^{\lambda\lambda}(p_1,p_2)$ is symmetric in $p_1,p_2$.

A useful test is that $\varphi_l(l_1)V\varphi_r(l_2)$ has to be Weyl invariant, keeping terms to the appropriate order. From $\delta_\sigma h_{\mu\nu}=-2\sigma \eta_{\mu\nu}$, $\delta_\sigma h=-8\sigma$, $\delta_\sigma \varphi_{l,r}=\sigma \varphi_{l,r}$ and $\delta_\sigma \lambda=\sigma$ one can derive in momentum space
\beq
\delta_\sigma (\lambda(q)V^\lambda(q)+h(q)V^{h}_{\mu\mu}(p_h)) \overset{!}=-l_2l_2 e^{u\, l_2l_2}-l_1l_1 e^{u\, l_1l_1}
\eeq
and
\beq
\delta_\sigma (\lambda(p_1)\lambda(q)V^{\lambda\lambda}(p_1,q)+h(q)\lambda(p_1)V^{h\lambda}_{\mu\mu}(q,p_1))_{{\cal O}(\lambda)} \overset{!}=-\lambda(p_1)\left(V^\lambda(p_1)_{l_1=-l_2-p_1}+V^\lambda(p_1)_{l_2=-l_1-p_1}\right)
\eeq
where $q$ is identified with the momentum of the Weyl mode $\sigma$. (The two prescriptions in parenthesis on the right hand side are not equivalent since momentum conservation alone gives $l_1+l_2+p_1+q=0$. These prescriptions indicate that $q$ disappears from the right hand side if expressed in terms of the momenta of $\varphi_{l,r}$.)

We have verified that these relations are satisfied for our explicit expressions for the components of $V$. Below we give the ones for $V^\lambda(p_1)$ and $V^h_{\mu\nu}(p_h)$; those for $V^{\lambda\lambda}(p_1,p_2)$ and $V^{h\lambda}_{\mu\nu}(p_h,p_1)$ are considerably longer.

\begin{align}
V^\lambda(p_1)&=
-\frac{1}{12}\big\{(8(l_2 p_1)^3 + 12(l_2 p_1)^2(p_1 p_1) + 6(l_2 p_1)(p_1 p_1)^2 +(p_1 p_1)^3)u^3\phantom{xxxxxxxxxxxxxxxxxxx} \nn \\
& + (16(l_2 p_1)^2 + 16(l_2 p_1)(p_1 p_1) + 4(p_1 p_1)^2)u^2 + (24(l_2 p_1) + 12(p_1 p_1))u\nn \\
& + 24\big\}(l_1 l_1)(l_2 l_2)u
\end{align}

%\newpage
%\bea
\begin{align}
V^h_{\mu\nu}(p_h)&= \Big\{
\frac{1}{144}(24(l_1 l_1)(l_2 p_h)^4 + 44(l_1 l_1)(l_2 p_h)^3(p_h p_h) +30(l_1 l_1)(l_2 p_h)^2(p_h p_h)^2 \nn \\ &+
 9(l_1 l_1)(l_2 p_h)(p_h p_h)^3 + (l_1 l_1)(p_h p_h)^4)u^4 \nn \\ &+
\frac{1}{36}(12(l_1 l_1)(l_2 p_h)^3 + 16(l_1 l_1)(l_2 p_h)^2(p_h p_h) + 7(l_1 l_1)(l_2 p_h)(p_h p_h)^2 +(l_1 l_1)(p_h p_h)^3)u^3 \nn \\ &+
\frac{1}{12}(6(l_1 l_1)(l_2 p_h)^2 + 5(l_1 l_1)(l_2 p_h)(p_h p_h) +(l_1 l_1)(p_h p_h)^2)u^2 \nn \\  &+
\frac{1}{6}(3(l_1 l_1)(l_2 p_h) + (l_1 l_1)(p_h p_h))u + \frac{1}{2}(l_1 l_1) + \frac{1}{2}(l_1 p_h)+ \frac{1}{6}(p_h p_h)
\Big\} -l_{1\mu}l_{1\nu}  \nn \\
&+\Big\{-\frac{1}{24}(8(l_1 l_1)(l_2 p_h)^3 + 12(l_1 l_1)(l_2 p_h)^2(p_h p_h) + 6(l_1 l_1)(l_2 p_h)(p_h p_h)^2
+ (l_1 l_1)(p_h p_h)^3)u^4\nn \\ &-  
\frac{1}{6}(4(l_1 l_1)(l_2 p_h)^2 + 4(l_1 l_1)(l_2 p_h)(p_h p_h) +
(l_1 l_1)(p_h p_h)^2)u^3 \nn \\ &-   \frac{1}{2}(2(l_1 l_1)(l_2 p_h) + (l_1 l_1)(p_h p_h))u^2 - (l_1 l_1)u \Big\} l_{2\mu}l_{2\nu}\nn \\  &+
\Big\{-\frac{1}{144}(8(l_1 l_1)(l_2 p_h)^3 + 12(l_1 l_1)(l_2 p_h)^2(p_h p_h) +
6(l_1 l_1)(l_2 p_h)(p_h p_h)^2 + (l_1 l_1)(p_h p_h)^3)u^4  \nn \\ &-
\frac{1}{36}(4(l_1 l_1)(l_2 p_h)^24(l_1 l_1)(l_2 p_h)(p_h p_h) + (l_1 l_1)(p_h p_h)^2)u^3  \nn \\ &- 
\frac{1}{12}(2(l_1 l_1)(l_2 p_h) + (l_1 l_1)(p_h p_h))u^2
- \frac{1}{6}(l_1 l_1)u - \frac{1}{6}\Big\} p_{h\mu}p_{h\nu} - p_{1\mu}p_{h\nu}  \nn \\  &+
\Big\{-\frac{1}{24}(8(l_1 l_1)(l_2 p_h)^3 + 12(l_1 l_1)(l_2 p_h)^2(p_h p_h) + 6(l_1 l_1)(l_2 p_h)(p_h p_h)^2
+ (l_1 l_1)(p_h p_h)^3)u^4 \nn \\ &-
 \frac{1}{6}(4(l_1 l_1)(l_2 p_h)^2 + 4(l_1 l_1)(l_2 p_h)(p_h p_h) +(l_1 l_1)(p_h p_h)^2)u^3 \nn \\ &-
 \frac{1}{2}(2(l_1 l_1)(l_2 p_h) + (l_1 l_1)(p_h p_h))u^2 - (l_1 l_1)u \Big\} p_{2\mu}p_{h\nu}\; .
%\eea
\label{eq:Vhmunu}
\end{align}

The term $\sqrt{\gamma}\, m_0^2\, \varphi^2$ in eq.~\eqref{eq:skm} expanded in $h_{\mu\nu}$ generates vertices proportional to $m_0^2\, \varphi^2$ given in eq.~\eqref{eq:cts}, and proportional to $h\, m_0^2\, \varphi^2$ with $m_0^2$ as in eq.~\eqref{eq:cts}. However, the latter vertices do not contribute to ${\cal G}_{\chi_i\chi_j}$ obtained according to eq.~\eqref{eq:GAFE}. The only one-loop diagram which contributes to ${\cal G}^0_{\nu\nu}$ in eq.~\eqref{eq:gnunu0} contains two vertices $V^\nu = 2\overline{m}^2$, and one vertex $V^h_{\mu\nu}$ given in eq.~\eqref{eq:Vhmunu}.

To ${\cal O}(g)$, two-loop diagrams which contribute to $\Delta {\cal G}^0_{\nu\nu}$ consist in vertex-less tadpoles attached to the one-loop diagram with two vertices $V^\nu$ and one vertex $V^h_{\mu\nu}$, and a tadpole containing one vertex $V^\nu$ attached to the one-loop diagram with one vertex $V^\nu$ and one vertex $V^h_{\mu\nu}$. Quadratic divergences are cancelled by one-loop diagrams where the tadpoles are replaced by corresponding counter terms from $m_0^2$. Altogether finite contributions for $\overline{\Lambda}^2\to \infty$ cancel as well which goes beyond the required cancellation of subdivergences.

Diagrams which contribute to $\Delta {\cal G}^0_{\nu\lambda}$ consist in a loop with vertices $V^h_{\mu\nu}$ and $V^\lambda$ with a tadpole containing one vertex $V^\nu$ together with a corresponding counter term, and a loop with vertices $V^h_{\mu\nu}$ and $V^\nu$ with a tadpole containing one vertex $V^\lambda$ together with a corresponding counter term. 
%All contributions are suppressed for $\overline{\Lambda}^2\to \infty$. The contributions to $\Delta {\cal G}^0_{\lambda\lambda}$ can be deduced from eqs.~\eqref{eq:cons1}.

\end{document}